\documentclass[namedreferences]{solarphysics}
\usepackage[hyperref,optionalrh,solaromanenum]{spr-sola-addons} 
\usepackage{graphicx}                    
\usepackage{color}                       

\chardef\us=`\_

\usepackage{siunitx} 
\usepackage{soul} 

\begin{document}

\begin{article}

\begin{opening}

\title{On-disc Observations of Flux Rope Formation Prior to its Eruption}

%

\author[addressref={1},corref,email={alexander.james.15@ucl.ac.uk}]{\inits{A.W. }\fnm{A.W. }\lnm{James}~\orcid{0000-0001-7927-9291}}

\author[addressref={1},corref,email={}]{\inits{L.M. }\fnm{L.M. }\lnm{Green}~\orcid{0000-0002-0053-4876}}

\author[addressref={2},corref,email={}]{\inits{E. }\fnm{E. }\lnm{Palmerio}~\orcid{0000-0001-6590-3479}}

\author[addressref={1},corref,email={}]{\inits{G. }\fnm{G. }\lnm{Valori}~\orcid{0000-0001-7809-0067}}

\author[addressref={3},corref,email={}]{\inits{H. }\fnm{H.A.S. }\lnm{Reid}~\orcid{0000-0002-6287-3494}}

\author[addressref={1},corref,email={}]{\inits{D. }\fnm{D. }\lnm{Baker}~\orcid{0000-0002-0665-2355}}

\author[addressref={4},corref,email={}]{\inits{D.H. }\fnm{D.H. }\lnm{Brooks}~\orcid{0000-0002-2189-9313}}

\author[addressref={1,5,6},corref,email={}]{\inits{L. }\fnm{L. }\lnm{van Driel-Gesztelyi}~\orcid{0000-0002-2943-5978}}

\author[addressref={2},corref,email={}]{\inits{E. K. J. }\fnm{E.K.J.  }\lnm{Kilpua}~\orcid{0000-0002-4489-8073}}

%

\runningauthor{A.W. James \textit{et al.}}
\runningtitle{On-disc Observations of Flux Rope Formation Prior to its Eruption}


\address[id={1}]{University College London, Mullard Space Science Laboratory, Holmbury St. Mary, Dorking, Surrey,
RH5 6NT, UK}

\address[id={2}]{University of Helsinki, Department of Physics, P.O. Box 64, 00014 Helsinki, Finland}

\address[id={3}]{SUPA School of Physics and Astronomy, University of Glasgow, Glasgow, G12 8QQ, UK}

\address[id={4}]{College of Science, George Mason University, 4400 University Drive, Fairfax, VA 22030, USA}

\address[id={5}]{Observatoire de Paris, LESIA, FRE 2461 (CNRS), 92195 Meudon Principal Cedex, France}

\address[id={6}]{Konkoly Observatory of the Hungarian Academy of Sciences, Budapest, Hungary}



\begin{abstract}
	Coronal mass ejections (CMEs) are one of the primary manifestations of solar activity and can drive severe space weather effects. Therefore, it is vital to work towards being able to predict their occurrence. However, many aspects of CME formation and eruption remain unclear, including whether magnetic flux ropes are present before the onset of eruption and the key mechanisms that cause CMEs to occur. In this work, the pre-eruptive coronal configuration of an active region that produced an interplanetary CME with a clear magnetic flux rope structure at 1 AU is studied. A forward-S sigmoid appears in extreme-ultraviolet (EUV) data two hours before the onset of the eruption (SOL2012-06-14), which is interpreted as a signature of a right-handed flux rope that formed prior to the eruption. Flare ribbons and EUV dimmings are used to infer the locations of the flux rope footpoints. These locations, together with observations of the global magnetic flux distribution, indicate that an interaction between newly emerged magnetic flux and pre-existing sunspot field in the days prior to the eruption may have enabled the coronal flux rope to form via tether-cutting-like reconnection. Composition analysis suggests that the flux rope had a coronal plasma composition, supporting our interpretation that the flux rope formed via magnetic reconnection in the corona. Once formed, the flux rope remained stable for 2 hours before erupting as a CME.
\end{abstract}

%


\end{opening}


%

%

%

\section{Introduction}\label{Intro} 

Coronal mass ejections (CMEs) are eruptions of billions of tonnes of plasma from the Sun, requiring large amounts of energy ($\approx$10$^{32}$ ergs). The only coronal energy source large enough to produce CMEs is magnetic energy, as kinetic, gravitational and thermal energy fall short by orders of magnitude \citep{forbes2000review}. While it has been theorised that CMEs may be a method of releasing magnetic helicity in addition to magnetic energy from the corona \citep{rust1994spawning}, the physical processes behind the initiation and subsequent evolution of the eruptions remain unclear.

CMEs that are able to escape the Sun and propagate through the heliosphere are sometimes referred to as interplanetary CMEs (ICMEs) when they are detected \textit{in situ}. When ICMEs are Earth-directed, they can drive phenomena that affect the Earth and the near-Earth environment, and these effects are known collectively as space weather. Space weather impacts include geomagnetically induced currents that can cause strong voltage fluctuations and heating in power grid transformers, and solar energetic particles that can damage satellites. Therefore, early space weather forecasts are needed to allow preparation for the effects of Earth-directed ICMEs.

Spacecraft measuring the solar wind \textit{in situ} detect signatures of ICMEs. ICMEs appear as strong magnetic field regions that sometimes display a smooth rotation in field direction as the ICME passes the spacecraft. Such observations are called ``magnetic clouds'' \citep{burlaga1981magnetic} and are consistent with the presence of a flux rope: a twisted magnetic structure featuring field lines wrapped helically around an axis. 

It is important to further our understanding of the pre-eruptive coronal magnetic field configuration of CMEs so we can determine what physical processes cause them to occur. Currently, we have to wait until a CME is in progress before space weather forecasting can begin, with measurements of the ICME properties being taken \textit{in situ} between the Sun and the Earth at the first Lagrange point (L1). A moderate CME speed at L1 of 400 km s$^{-1}$ (as in \citealp{gopalswamy2000interplanetary}) implies a typical propagation time of around an hour from L1 to Earth in which to prepare for the arrival of an ICME. Gaining the ability to predict whether an active region might produce a CME could improve lead-times of space weather forecasts by a matter of days.

There are two groups of theories about the pre-eruptive magnetic field configuration \citep{forbes2000review,forbes2006CME,aulanier2010formation}. One of these involves the presence of a sheared arcade. Sheared arcades are non-potential and are therefore a store of magnetic energy that may be able to power an eruption. In these sheared arcade eruption models, a flux rope may form during the onset of eruption (\textit{e.g.}, via tether-cutting reconnection; \citealp{moore2001onset}) or after the onset of eruption (\textit{e.g.}, via magnetic breakout; \citealp{antiochos1999model,lynch2008topological}). The other group of theories requires that the pre-eruptive configuration is that of a flux rope. In this work, a flux rope will be defined as a structure that features an axial field line with at least one off-axial poloidal magnetic field line wrapped around it with at least one full turn (2$\pi{}$) from end-to-end. This is the same as outlined by \cite{savcheva2012photospheric}. Field lines that feature less twist than this are classified as belonging to a sheared arcade.

A flux rope that is present before an eruption could be formed by a series of reconnection events low down in the solar atmosphere. For example, \citet{van1989formation} described how the shearing, convergence and reconnection of magnetic field driven by photospheric flows may form a flux rope that could contain filament material. A decrease in magnetic flux is observed in the photosphere when resultant low-lying small-scale reconnected loops submerge beneath the photosphere due to magnetic tension. This process is therefore called flux cancellation. A flux rope formed by flux cancellation should form low down with its underside in the photosphere or chromosphere due to the low altitude of the reconnection. Alternatively, pre-eruption flux ropes can also form via reconnection in the corona. Photospheric motions are likely to also be responsible for their formation, but in this case by bringing together coronal field lines that then reconnect.

A distinguishing factor between the mechanisms of flux rope formation is the composition of plasma the flux ropes contain. A flux rope that forms as a result of coronal reconnection (as seen in \citealp{patsourakos2013direct}) would contain coronal plasma, in contrast to the low-lying \citet{van1989formation} process that feeds photospheric or chromospheric plasma into the flux rope. Plasma composition can be determined by studying the intensity of emission lines from elements with different first ionisation potentials (FIP; \citealp{brooks2011establishing}). Relative to the photosphere, coronal plasma has a greater abundance of low-FIP elements. \citet{baker2013plasma} observed photospheric plasma in the core of a sigmoidal active region, and suggested this could correspond to part of a flux rope that formed via reconnection low-down in the solar atmosphere via flux cancellation along the polarity inversion line. In addition, the flux cancellation associated with the \citet{van1989formation} model is observed to occur on timescales of days, but the rate of formation of coronal flux ropes is harder to quantify, largely because they are most commonly observed at the limb where they are easier to see \citep{Nindos2015common}. The reconnection that builds the flux rope could be an ongoing, gradual process, or it could occur in bursts caused by events in the local active region. The occurrence of magnetic reconnection can be probed with radio data because energetic electrons accelerated along magnetic field lines in the reconnection outflow can emit at radio wavelengths. EUV and soft X-ray data may also show increases in intensity due to magnetic reconnection. Some studies have observed signatures of hot coronal flux ropes that erupt $\approx$10 minutes after they first brighten \citep{cheng2011observing,zhang2012observation}, whereas \citet{patsourakos2013direct} observed a flux rope appear and grow larger over 17 minutes that then erupted $\approx$7 hours later.

Pre-eruption flux ropes may then erupt as a result of ideal or non-ideal processes. Ideal instabilities include the kink instability, which occurs when the twist of the erupting structure exceeds a critical value \citep{torok2005confined}, and the torus instability, which occurs when there is a rapid drop-off in vertical magnetic field strength above an arched flux rope \citep{kliem2006torus}. Non-ideal CME models involve magnetic reconnection (\textit{e.g.}, tether-cutting reconnection; \citealp{moore2001onset}, and magnetic breakout; \citealp{antiochos1999model,lynch2008topological}).

Unfortunately, there are difficulties in directly measuring the coronal magnetic field configuration. Measurements of the photospheric magnetic field can be produced by exploiting the Zeeman effect, but because temperatures in the corona can be more than 1000 times larger than in the photosphere \citep{gary2001plasma}, thermal and non-thermal broadening have a much larger effect than Zeeman splitting in the corona. Furthermore, because the corona is optically thin, it is difficult to determine the altitude from which the observed photons originate, and various magnetic field directions and strengths may occur along the line of sight.

However, the structure of the corona can be inferred without directly measuring the coronal magnetic field. For example, an inverse crossing in vector photospheric magnetic field data refers to a transverse field component that crosses a polarity inversion line from negative to positive; the opposite way to how a potential field would behave, and may correspond to a ``bald patch'' of magnetic field \citep{titov1993conditions}. An inverse crossing is a necessary (but not sufficient) signature of a low-lying flux rope with helical field that intersects the photosphere. There can also be coronal signatures of flux ropes, such as sigmoids and plasmoids. Sigmoids are continuous S-shaped (or inverse-S-shaped) plasma structures that appear strongly in X-ray bands, but also at hot extreme-ultraviolet (EUV) wavelengths. Their shape is representative of weakly twisted magnetic field bundles and they may be an indicator of flux rope presence \citep{rust1996evidence,green2009flux}. Plasmoids are generally features seen in events near the solar limb. The term describes hot, globular plasma structures that may be interpreted as a cross-section of a flux rope as it is viewed along its axis (for examples, see \citealp{shibata1995hot,reeves2011atmospheric}). Furthermore, measurements of magnetic clouds \textit{in situ} allow the linking of magnetic structures detected in the solar wind back to their point of origin at the Sun \citep{marubashi1986structure,mcallister2001prediction,yurchyshyn2001orientation,mostl2008reconstruction,yurchyshyn2008relationship}. These measurements can be used alongside solar observations to gain a more complete picture of the eruptive event. 

In this work, the solar origin of a magnetic cloud is investigated to determine its pre-eruptive configuration. EUV observations from SDO/AIA are used to study the evolution of NOAA active region 11504 before, during and after it produced a CME on 14 June 2012. These are supplemented with photospheric line-of-sight and vector magnetic field measurements from SDO/HMI, as well as spectroscopic measurements from \textit{Hinode}/EIS and radio observations from the \textit{Nan\c{c}ay Radioheliograph}. The various instruments and data products used are outlined in Section \ref{Instruments}. Section \ref{Observations} begins by describing the \textit{in situ} magnetic cloud detection and how the active region of origin is identified. The magnetic field, EUV, spectroscopic and radio observations of the active region are also presented in Section \ref{Observations}, and their implications are discussed in Section \ref{Discussion}. Finally, the key conclusions are given in Section \ref{Conclusions}. A study of the same magnetic cloud was performed by \citet{palmerio2017insitu}, in which the compatibility between the \textit{in situ} measurements and the pre-eruptive configuration was examined.

\section{Data Sources and Processing} \label{Instruments}

Coronagraph images were taken using the C2 coronagraph of the \textit{Large Angle and Spectrometric Coronagraph} instrument (LASCO; \citealp{brueckner1995large}) onboard the \textit{Solar and Heliospheric Observatory} (SOHO; \citealp{domingo1995soho}) to identify the solar eruption that produced the magnetic cloud. LASCO also provides an estimate of plane-of-sky CME speed soon after the eruption.

\overfullrule=0pt 
Two instruments from the \textit{Solar Dynamics Observatory} (SDO; \citealp{pesnell2012SDO}) are used for low-coronal and photospheric characterisation. The \textit{Atmospheric Imaging Assembly} (AIA; \citealp{lemen2012atmospheric}) produces full-disc EUV images at seven wavelengths with a cadence of 12 seconds and  pixels that correspond to angular diameters of 0.6$''$ on the Sun. The aia\_{}prep routine in SolarSoft is used to prepare the AIA data to level 1.5, correcting for the slight difference in viewing angle, focal length and alignment between each of the four AIA telescopes.

The \textit{Helioseismic and Magnetic Imager} (HMI; \citealp{scherrer2012helioseismic}) has two cameras with pixel sizes of 0.5$''$. One produces line-of-sight magnetograms and white light continuum images at a cadence of 45 seconds, and the other makes vector magnetic field measurements at an overall cadence of 12 minutes. The vector dataset used in this work is the hmi.sharp\_cea\_720s series, which provides automatically-selected cutouts of active regions (space weather HMI active region patches; SHARPs --- see \citealp{bobra2014helioseismic}) in the form of Lambert cylindrical equal-area (CEA) projection. These data have a minimum-energy optimisation process applied to them with the aim of solving the 180$^{\circ}$ ambiguity in the azimuthal magnetic field direction. HMI SHARP data were not accessible during 13 June 2012 (the day before the eruption), so instead, cutouts are taken from the full-disk vector magnetic field series and transformed into the CEA projection. These were then hand-aligned with the SHARP CEA data. The line-of-sight data are processed using the aia\_prep routine to change the image scale to 0.6$''$ per pixel to match AIA for easier comparison between the two data sets.

The \textit{Geosynchronous Operational Environmental Satellite} (GOES) system is used to provide a full-disc X-ray flux lightcurve over the hours before, during and after the eruption, as well as a list of flares that were observed in the active region during this time.
 
Spectroscopic measurements were made using the \textit{EUV Imaging Spectrometer} (EIS; \citealp{culhane2007euv}) onboard \textit{Hinode}. The data we use are 1$''$-slit coarse (2$''$) raster scans of a 120$''\times$512$''$ field-of-view, taking 60 second exposures at each position. They are processed and calibrated using standard routines available in SolarSoft. These procedures remove the dark current pedestal, account for hot, warm, and dusty pixels, and correct the orbital drift of the spectrum on the CCD. The data are then calibrated to physical units. The EIS calibration shows an on-orbit evolution with time \citep{del2013revised,warren2014absolute}, so we also re-calibrate the data using the method of \citet{del2013revised}. The EIS scans used here are manually aligned with AIA images to aid comparison of features.

Finally, we use radio images made by the \textit{Nan\c{c}ay Radioheliograph} (NRH; \citealp{kerdraon1997nanccay}) at nine frequencies between 150 MHz and 445 MHz using a 10 second integration time.

\section{Observations} \label{Observations}
\subsection{The Magnetic Cloud \textit{in situ} and CME Association} \label{in_situ}
An \textit{in situ} observation of a magnetic cloud was made by the \textit{Wind} satellite on 16 June 2012. The shock preceding the magnetic cloud was first detected at $\approx$19:30 UT on 16 June, and the passage of the magnetic cloud lasted from $\approx$22:00 UT on 16 June until $\approx$12:30 UT on 17 June. The measured magnetic field vector rotated from north to south in the geocentric solar ecliptic (GSE) coordinate system as the cloud passed the spacecraft, and an eastern field component was measured throughout the cloud (see Figure 6 of \citealp{palmerio2017insitu}). This is consistent with a flux rope that has eastern axial field with helical field wrapped around it. The helical field at the leading edge of the flux rope is northward and at the trailing edge it is southward, so the measured rotation from north to south is produced as it passes over the spacecraft. The \textit{in situ} flux rope detection is discussed in more detail by \citet{palmerio2017insitu} in addition to another event, whereas in this paper we focus on the pre-eruptive coronal field that generated the magnetic cloud.

The magnetic cloud has previously been associated with a halo CME that was observed by LASCO on 14 June 2012 at 14:12 UT (Richardson-Cane ``Near-Earth Interplanetary Coronal Mass Ejections'' list;\footnote{\url{http://www.srl.caltech.edu/ACE/ASC/DATA/level3/icmetable2.htm}} \citealp{Kubicka2016prediction}; \citealp{palmerio2017insitu}. For the method used in creating the ICME list, see \citealp{Cane2003interplanetary} and \citealp{Richardson2010near}). Here we check the validity of the association. Images from the STEREO A and B satellites confirm that the halo CME occurred from the Earth-facing side of the Sun, and the ICME speed was measured to be $\approx$500 km s$^{-1}$ by \textit{Wind} at 1 AU. Assuming the ICME had a constant speed between the Sun and L1 suggests the source eruption took place at $\approx$08:30 UT on 13 June 2012. However, since the speed of the solar wind preceding the ICME was lower than the ICME speed (solar wind speed $\approx$400 km s$^{-1}$ measured by \textit{Wind}), it is likely the ICME had been decelerating before reaching L1. This means the time of 08:30 UT on 13 June is an earliest limit on the eruption time and the eruption likely took place somewhat later than this. 

There are six eruptions listed in the ICME catalog\footnote{See \url{https://cdaw.gsfc.nasa.gov/CME_list/.}} that occurred between this earliest limit of 08:30 UT on 13 June and the halo eruption on 14 June. Halo eruptions manifest when CMEs travel along the line-of-sight of an observer, and so it is these eruptions that are most likely to result in Earth-directed CMEs. One of the six other eruptions was a partial halo CME, but that eruption is associated with a small, low-speed CME that was detected by \textit{Wind} 12-14 hours prior to the magnetic cloud referenced in this work. This association is made because the 13 June partial halo eruption was determined to have a slower speed than the full halo eruption of the 14th by SOHO LASCO. This therefore leads to the association of the 14 June halo CME with the 16 June magnetic cloud.

On 14 June 2012, a halo CME was observed in LASCO C2 coronagraph images at 14:12 UT. The radius of the occulting disc used in the C2 coronagraph is 2 R\textsubscript{$\odot$} ($\approx$1.4$\times10^{6}$ km), so combining this information with the linear plane-of-sky speed of $\approx$980 km s$^{-1}$ determined by LASCO suggests an eruption on the Sun approximately 20--30 minutes before its detection. The full-disc EUV data from 13:00 UT--14:00 UT on 14 June 2012 show that an eruption originated from NOAA active region 11504 during this time (see Section \ref{Obs_Cor}), and therefore we associate the origin of the considered ICME with NOAA active region 11504.

\subsection{Photospheric Evolution} \label{HMI_evolution}
 \begin{figure} 
 \centerline{\includegraphics[width=0.96\textwidth,clip=]{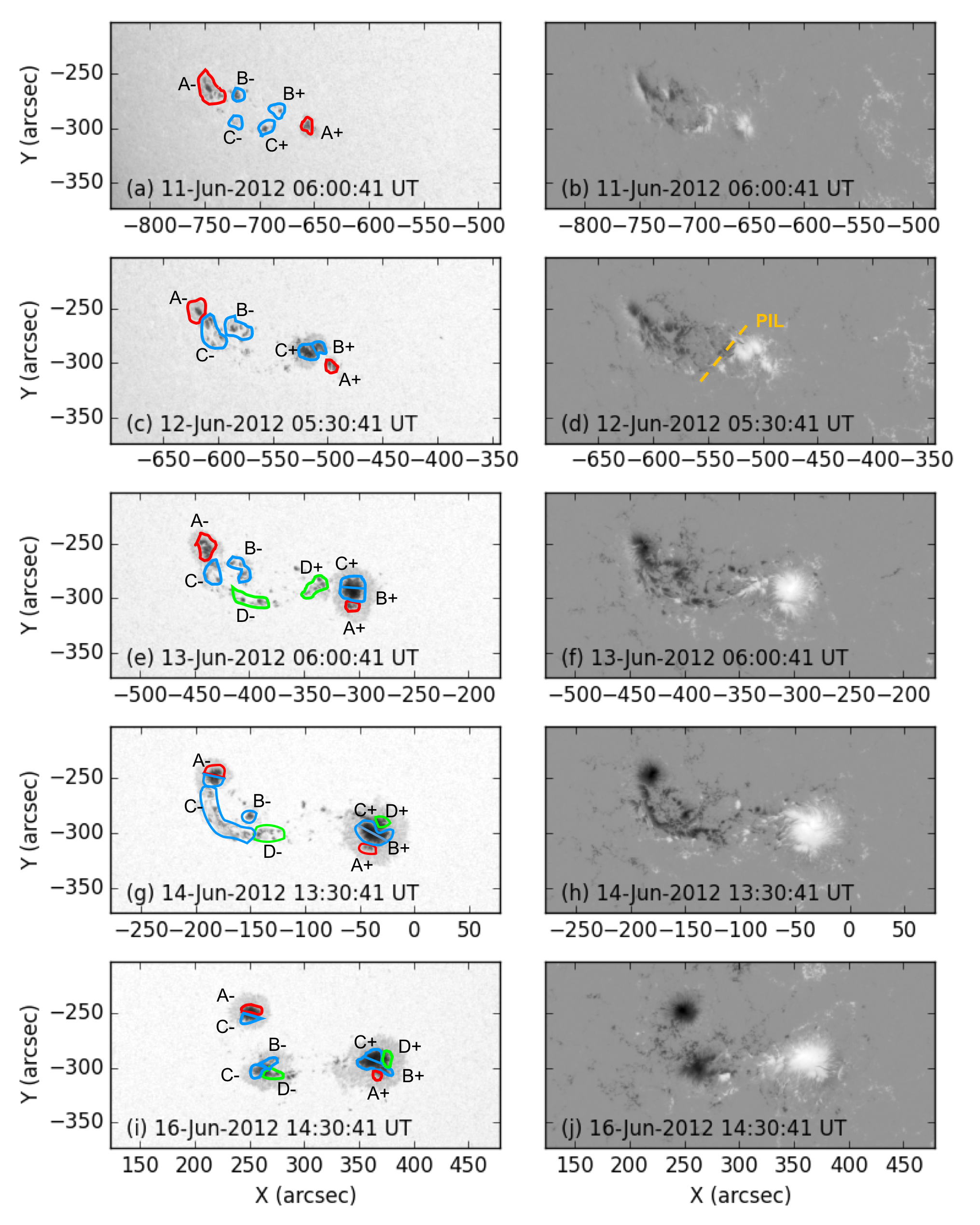}}
 \caption{Annotated continuum intensity images (left) and line-of-sight magnetograms (right) from HMI showing the evolution of NOAA active region 11504. The magnetograms are saturated to $\pm$2000 G, with positive flux shown in white and negative in black. (a/b) The region contains two small sunspots (red, A$\pm$) as it rotates onto the disc. (c/d) Flux emergence (blue, B$\pm$, C$\pm$) occurs between the two sunspots. The magnetic tongue configuration is demonstrated by the annotated polarity inversion line (PIL) in panel (d). (e/f) The first positive emergent flux (B+, C+) joins the western sunspot (A+) as a separated umbra within the penumbra, but the negative flux (C-) forms an elongated tongue. The southernmost emerged positive flux (C+) has moved northward and clockwise around the other (B+), swapping north-south positions. Flux emergence is still ongoing (green, D$\pm$). (g/h) This further flux emergence adds a third umbra to the western sunspot (D+) and builds the eastern magnetic tongue (D-). (i/j) In the days after the CME, most of the elongated negative flux forms a third sunspot (B-, part of C-, D-). The northernmost section of the western sunspot (D+) from panels (g/h) has moved clockwise around the sunspot to a western position.}
 \label{fig:HMI_multi}
 \end{figure}
 \begin{figure} 
 \centerline{\includegraphics[width=0.8\textwidth,clip=]{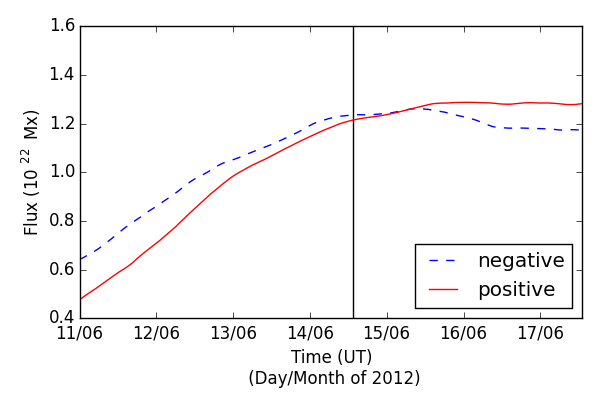}}
 \caption{Integrated magnetic flux of NOAA active region 11504 during its disk passage. Vertical field-component HMI SHARP images are smoothed over $15\times{}15$ pixel boxes and then only pixels with a magnetic field strength greater than $\pm{}$400 G are included in order to exclude the majority of quiet Sun field. This therefore fails to count some small-scale serpentine flux emergence, and so these values of magnetic flux should be regarded as lower limits. 24-hour moving-average smoothing has been applied to the curve to remove short time-scale variations. The vertical black line is drawn at 13:30 UT on 14 June to represent the time the eruption of the sigmoid is observed to begin.}
 \label{fig:Flux}
 \end{figure}
 \begin{figure} 
 \centerline{\includegraphics[width=1.0\textwidth,clip=]{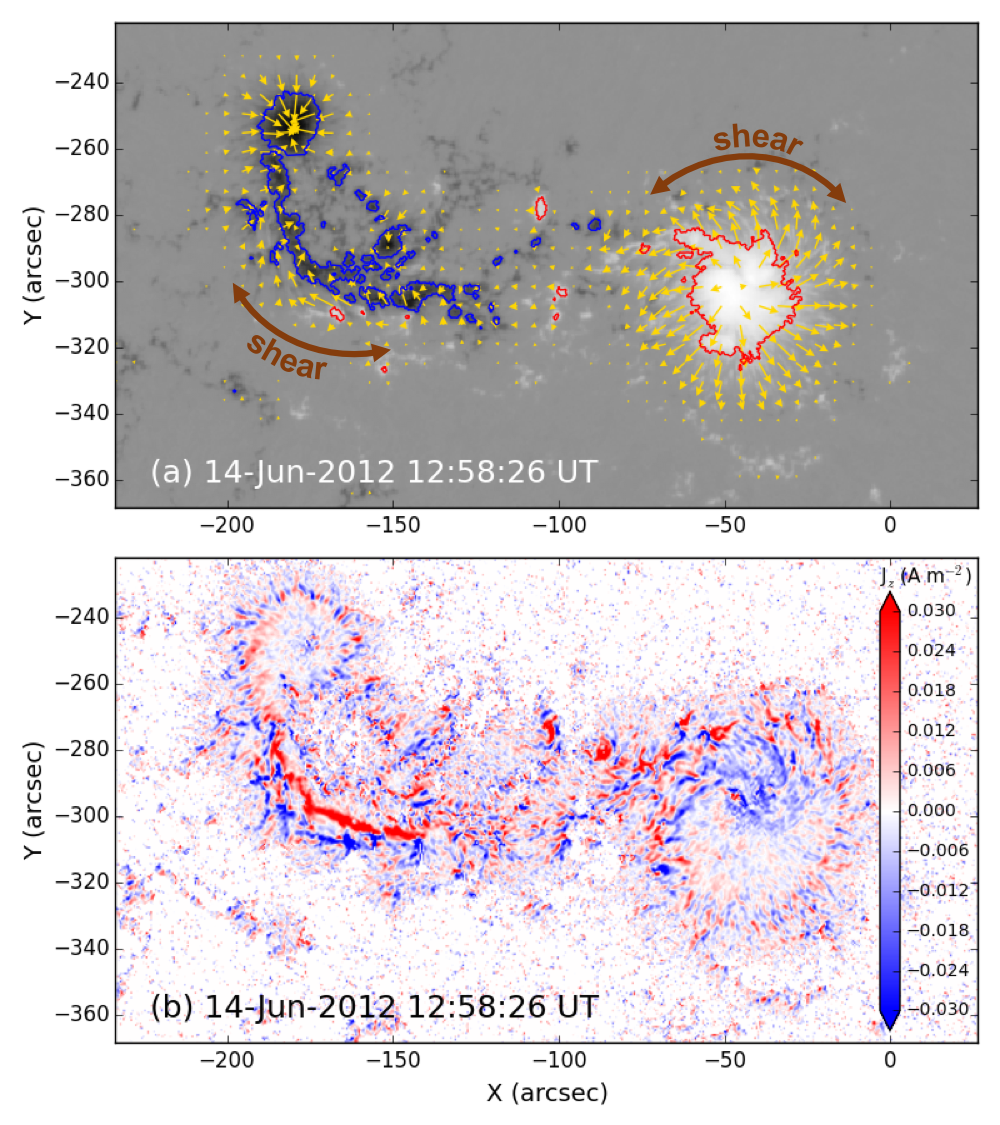}}
 \caption{(a): The vertical component of the vector magnetic field, with positive polarity in white and negative polarity in black. Red and blue contours outline regions of $\pm$1000 G. Gold arrows represent the direction and magnitude of the horizontal field component for horizontal field strengths of 100 G or higher. The horizontal field component is sheared to the north of the positive sunspot and in the negative magnetic tongue. (b): A map of vertical current density (\textit{J\textsubscript{z}}) in units of A m$^{-2}$  (red=positive, blue=negative) with the same field-of-view as panel (a), demonstrating locations of strong gradients in the horizontal magnetic field component. The calculation of current densities in these units relies on the assumption that each pixel represents an equal area in square-metres. This assumption breaks down the further away from disc-centre due to projection effects. To remove noise, we set \textit{J\textsubscript{z}}=0 in pixels where the horizontal field strength is weaker than 100 G.}
 \label{fig:Vector_Jz}
 \end{figure}

NOAA active region 11504 rotates onto the solar disc as seen by SDO on 8 June 2012 and rotates off on 21 June 2012. The region initially contains two small sunspots of leading positive polarity and trailing negative polarity (see Figure \ref{fig:HMI_multi}a/b). New magnetic flux emergence occurs between the two spots and they separate (see Figure \ref{fig:HMI_multi}c/d). Figure \ref{fig:Flux} shows that $\approx$\num{7e21} Mx cm$^{-2}$ (with an estimated error of $\approx$5\% based on the 2.3 Mx cm$^{-2}$ noise-per-pixel value of \citealp{liu2012comparison}) of magnetic flux emerges in the active region from 11 June until around noon on 15 June, at which time emergence ceases. 
  
The positive flux that emerges from 11--13 June moves towards the north of the positive sunspot and is distinguished as two sections of umbra in Figure \ref{fig:HMI_multi}. The southernmost section (C+ in Figure \ref{fig:HMI_multi}) moves northward and clockwise around the northern emerged umbra (B+ in Figure \ref{fig:HMI_multi}), such that the southernmost emerged positive umbra eventually comprises the northernmost part of the positive sunspot. However, rather than fully coalescing with the pre-existing positive sunspot umbra, the newly-emerged flux remains somewhat separate from the pre-existing flux, existing as distinct umbrae within one penumbra (see Figure \ref{fig:HMI_multi}e). Further emergence of positive flux from 13--15 June forms another separated area of umbral field within the positive sunspot penumbra to the north of the previous sections (see D+ in Figure \ref{fig:HMI_multi}g). As it approaches the positive sunspot, it is observed to orbit clockwise around the sunspot. This continues, displaying a $\approx$120$^{\circ}$ movement about the sunspot in 24 hours before stopping (see Figure \ref{fig:HMI_multi}i). This orbit of newly emerged flux around separate areas of magnetic flux seen in the photosphere suggests that regions of magnetic field in the corona may be wrapping around each other --- increasing stored energy and creating more favourable conditions for magnetic reconnection between the different magnetic flux regions. The implications of this are discussed in full in Section \ref{Discussion}. The negative flux that emerges between 11 June and 15 June forms a negative magnetic tongue to the south of the negative sunspot (B-, C-, D- in Figure \ref{fig:HMI_multi}g). Magnetic tongues are a signature of the emergence of twisted flux through the photosphere and can therefore be used to observationally determine the chirality of emerging flux \citep{fuentes2000counterkink,luoni2011twisted}. In this active region, the negative tongue extends to the north of the positive emerging flux (see Figure \ref{fig:HMI_multi}d), indicating that the emerging flux tube has right-handed twist. The sheared negative tongue develops into a third sunspot, becoming cohesive on 16 June (see Figure \ref{fig:HMI_multi}i/j). From 15 June until the active region leaves the disc, the positive sunspot moves eastward towards the third sunspot. This motion is observed in the line-of-sight data, but also in the cylindrical equal-area projected vector data that remove some of the foreshortening near the limb. This suggests that the convergence is not just a foreshortening effect near the limb. Additionally, when the converging motion begins on 15 June, the active region is still quite far from the western limb, with the leading sunspot only $\approx$250$''$ west of central meridian.
 
The horizontal component of the magnetic field extends radially from the southern section of the positive sunspot (A+ in Figure \ref{fig:HMI_multi}), but the field extending out to its north (C+, D+) is sheared by $\approx$45$^{\circ{}}$ relative to the radial direction (see Figure \ref{fig:Vector_Jz}a). Furthermore, the pre-existing negative sunspot (A-) exhibits horizontal field that is mostly radial outwards from the sunspot, but the elongated tongue to its south (C-, D-) features shear to a similar degree as seen to the north of the positive sunspot. The positive and negative magnetic flux locations that exhibit horizontal field components without shear (A$\pm$) both pre-exist the flux emergence that is observed here, and the positive and negative field areas that do show shear in their horizontal field components emerge at around the same time as their opposing polarity counterparts (C$\pm$, D$\pm$). The negative tongue is associated with a region of relatively strong vertical electric current density, seen in Figure \ref{fig:Vector_Jz}b, and the same is seen to the north of the positive sunspot. The vertical current density is proportional to gradients in the horizontal field component, and so the vertical current is strongest in areas where shear is strongest. Finally, we find no inverse crossings along the central part of the polarity inversion line in the HMI vector data.
 
\subsection{Coronal Evolution} \label{Obs_Cor}

 \begin{figure}[t] 
 \centerline{\includegraphics[width=1.0\textwidth,clip=]{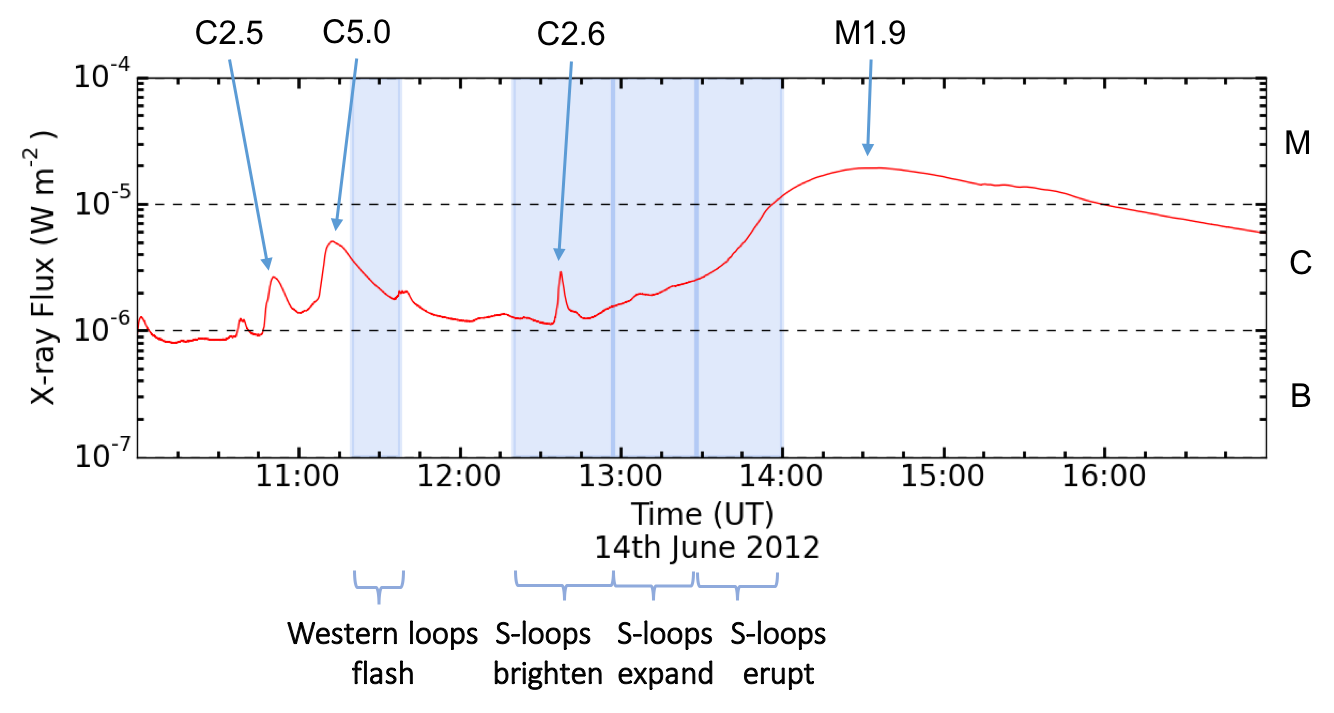}}
 \caption{The full-disc integrated X-ray flux (1.0-8.0 \AA{}) as measured by GOES-15. Flares observed in NOAA active region 11504 are labelled and the durations of observations from the 131 \AA{} EUV data described in Section \ref{Obs_Cor} are shown by the blue boxes. There is a two-stage rise in X-ray flux leading up to the CME, with a slow rise from $\sim$13:00 UT--13:30 UT, and a fast rise from 13:30 UT--14:00 UT.}
 \label{fig:GOES_timeline}
 \end{figure}
 \begin{figure} 
 \centerline{\includegraphics[width=1.0\textwidth,clip=]{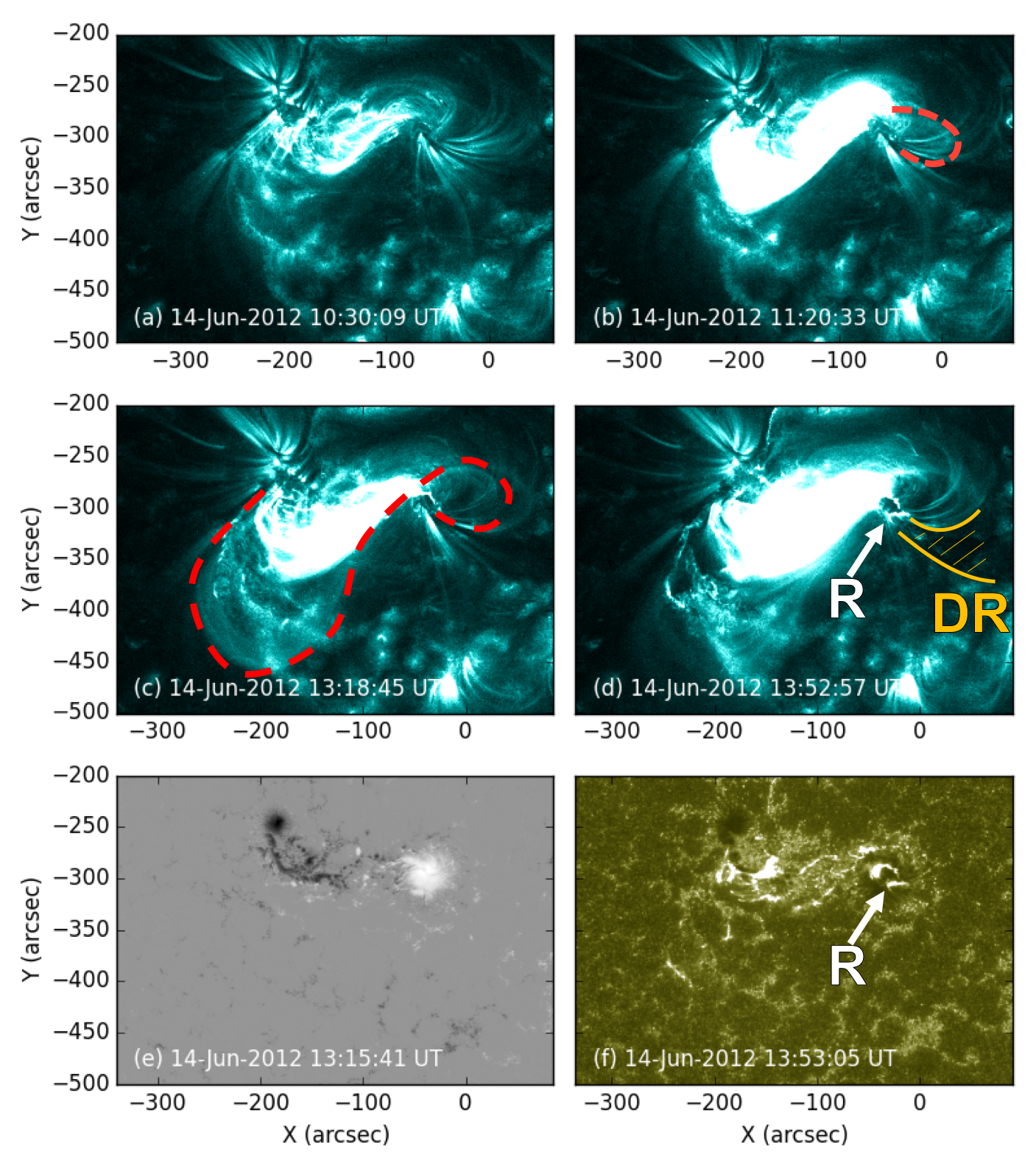}}
 \caption{Four EUV images at 131 \AA{} showing coronal evolution over the hours before, during and after the eruption, a line-of-sight magnetogram of the active region and an EUV image at 1600 \AA{} showing flare ribbons. (a) NOAA active region 11504 $\approx$3 hours before eruption. (b) Loops become visible to the west of the western sunspot (dashed line). (c) The full S shape is visible (dashed line). (d) The eruption is occurring, producing a hooked flare ribbon to the south of the western sunspot (annotated R). A dimming region (see Figure \ref{fig:Dimmings}a) that appears dark in EUV compared to surrounding plasma is outlined and hatched in gold, marked DR. (e) An HMI line-of-sight magnetogram of the active region. (f) The hooked flare ribbon (annotated R) is also visible at 1600 \AA{}. The AIA 131 \AA{} images are saturated to 200 DN s$^{-1}$ pixel$^{-1}$, the magnetogram is saturated to $\pm$2000 G and the AIA 1600 \AA{} image is saturated to 800 DN s$^{-1}$ pixel$^{-1}$.  
(An animated version of panels a-d of this figure is available in the online journal.)}
 \label{fig:131_Multi}
 \end{figure}
 \begin{figure}
 \centerline{\includegraphics[width=1.0\textwidth,clip=]{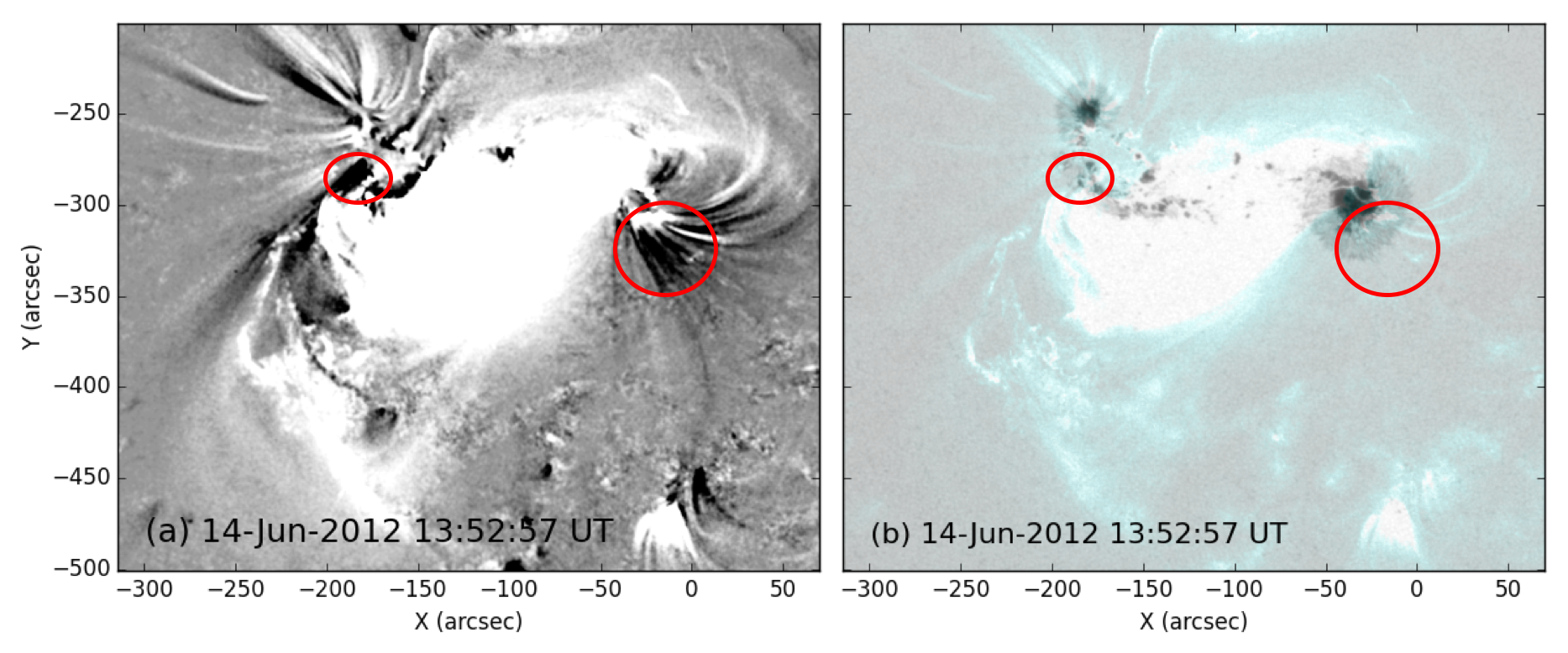}}
 \caption{(a) Base difference image of NOAA active region 11504 showing the difference in intensity between AIA 131 \AA{} images taken at 10:30 UT and 13:53 UT on 14 June 2012. Black areas correspond to decreases in intensity between the image times, and white represents brightenings (saturated to $\pm50$ DN s$^{-1}$ pixel$^{-1}$). The red circles indicate the areas near the east and west sunspots that undergo dimming around the time of eruption. (b) An HMI continuum white-light image of the active region with a semi-transparent 131 \AA{} AIA image overlayed. The AIA image is saturated to 500 DN s$^{-1}$ pixel$^{-1}$, as in Figure \ref{fig:131_Multi}. The dimming locations from panel (a) are included.}
 \label{fig:Dimmings}
 \end{figure}

Figure \ref{fig:GOES_timeline} depicts a timeline of coronal events observed in NOAA active region 11504 during the hours either side of the eruption on 14 June 2012. A C5.0 flare that produces a bright arcade in the centre of the active region begins at 11:05 UT and reaches its peak intensity at 11:12 UT (as measured by GOES). At 11:20 UT, loops appear to the west of the positive sunspot in the 131 \AA{} images of AIA (see Figure \ref{fig:131_Multi}b). Other AIA channels were studied, but the coronal features described in this section appear only in the 94 \AA{} channel and, most prominently, the 131 \AA{} channel ($\approx$6 MK and $\approx$10 MK respectively; \citealp{lemen2012atmospheric}). These western loops briefly brighten before fading out of view by 11:40 UT, by which time the intensity of the flare arcade has also decreased. Data from the \textit{X-ray Telescope} (XRT) onboard \textit{Hinode} were also studied, but there was a gap between observations from 11:12 UT - 11:43 UT.

At 12:20 UT, loops begin to appear to the south of the active region centre. A C2.6 flare begins at 12:33 UT and peaks at 12:37 UT. Throughout this time, the southern loops increase in brightness and size while remaining in the same location, and continue like this until 13:00 UT. As they brighten, loops once again appear to the west of the positive sunspot in the same location as those that appear at 11:20 UT, and it becomes clear that they are part of a common structure with the southern loops. In fact, together they form a continuous S-shape that traces along the polarity inversion line and has its ends rooted near each sunspot: a forward-S sigmoid (see Figure \ref{fig:131_Multi}c). This continuous sigmoid could not be identified in the XRT data because the field-of-view did not cover the whole active region.

At 12:52 UT, an M1.9 flare begins, although it does not reach peak intensity until 14:35 UT. The rise in X-ray intensity during this flare until its peak appears as a two-stage increase (see Figure \ref{fig:GOES_timeline}). From 12:52 UT until $\approx$13:30 UT, there is a relatively steady rise in intensity. During this time, the central flare arcade increases in brightness and the sigmoidal loops expand. However, from $\approx$13:30 UT until $\approx$14:00 UT, the X-ray intensity increases more rapidly. This coincides with a growth in the size and brightness of the flare arcade, as well as the eruption of the sigmoid, and the erupting sigmoid material is seen to move southward from the active region in the plane of the image. We conclude that the flare arcade forms beneath the sigmoid because the arcade is not disrupted by the upward motion and eruption of the sigmoid.

A flare ribbon is observed to brighten at the western footpoint of the flare arcade in every AIA channel from 13:45 UT. It is difficult to identify a flare ribbon on the eastern side due to the brightness of the arcade and the complex photospheric flux distribution, but the ribbon on the western side is particularly prominent, tracing around the group of positive sunspot umbrae. After extending out and around the sunspot, the flare ribbon turns back on itself to produce a hook-shape at 13:53 UT (see Figure \ref{fig:131_Multi}d).

At the same time as the flare ribbon on the western side traces out the hook, twin dimmings are observed in the EUV data, with one occurring on either side of the active region (marked with red circles in Figure \ref{fig:Dimmings}). These are most apparent when taking base difference images of the region, because these show increases and decreases in intensity relative to the chosen time of origin. The western dimming is located within the hook of the western ribbon (where the dark region labelled DR in Figure \ref{fig:131_Multi}d extends from), and the eastern dimming appears over a small patch of negative magnetic field to the south of the eastern sunspot. The locations of the dimmings and their significance to the pre-eruptive coronal configuration are discussed in Section \ref{Discussion}.

\subsection{Plasma Composition}

Composition analysis is performed using data from an EIS raster that began at 11:42 UT on 14 June 2012 to investigate whether features observed in the EUV data have photospheric or coronal plasma composition. We follow the method of \citet{brooks2011establishing}, which utilizes the Fe lines to measure the electron density and compute a differential emission measure (DEM) distribution. The density and temperature are then used to model the Si {\sc x} 258.375 \AA{} / S {\sc x} 264.223 \AA{} line ratio, which is sensitive to the difference in compositional fractionation of Silicon and Sulfur due to the FIP effect. The methodology has been tested extensively in many studies \citep{brooks2011establishing,brooks2012coronal,baker2013plasma,baker2015fip,culhane2014tracking,brooks2015observations,edwards2016comparison}, and we refer the reader to these works for specific examples. We also use the CHIANTI database v.8 to compute the contribution functions needed for this analysis \citep{dere1997chianti,del2015chianti}. 

\begin{figure} 
 \centerline{\includegraphics[width=1.0\textwidth,clip=]{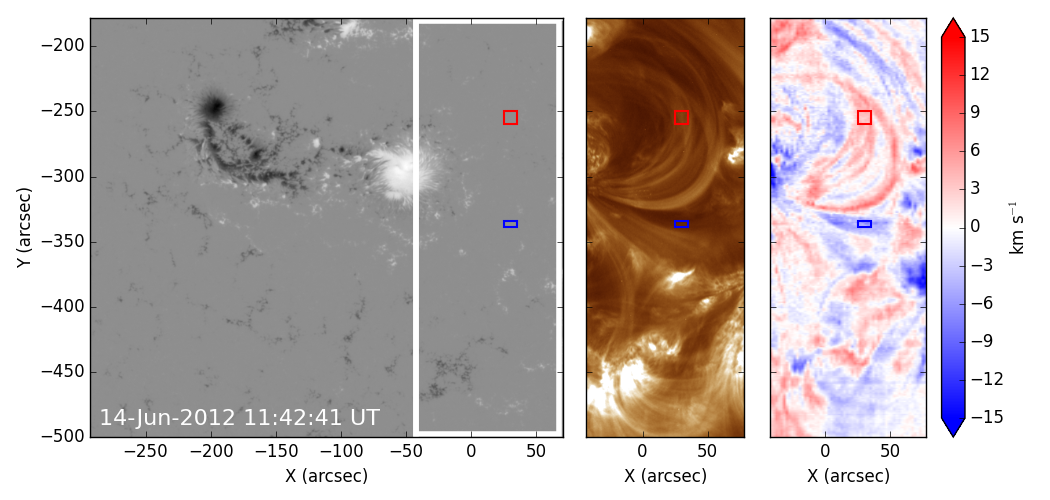}}
 \caption{HMI line-of-sight magnetogram (left), 193 \AA{} AIA image (middle) and an EIS Doppler velocity image produced at 195.12 \AA{} (Fe {\sc xii}, right). The AIA and EIS images show an area to the west of the positive sunspot, and this field-of-view is marked on the HMI image. Downflows (red) are observed by EIS in loops that extend to the north-west of the positive sunspot and upflows (blue) are observed to the south-west of the sunspot. The composition analysis is performed on these flow areas in locations shown by red and blue boxes respectively. Since these EIS images are produced via a raster that began at 11:42 UT, the red and blue boxes sample plasma that was observed at $\approx$12:00 UT. Doppler velocities are saturated to $\pm$15 km s$^{-1}$.}
 \label{fig:EIS_FIP}
 \end{figure}

The study is performed on loops that extend to the north-west of the positive sunspot, and also on an area to the south-west of the positive sunspot (see the red and blue boxes respectively in Figure \ref{fig:EIS_FIP}). These locations are chosen to probe plasma in the flux rope. The red box is located on a region that had been seen to form faint new loops in the AIA 131 \AA{} waveband around 40 minutes earlier (from 11:20 UT, as described previously, --- see Figure \ref{fig:131_Multi}b). We infer that these loops mark the location of the periphery of the forming flux rope, since the brightening is indicative of heating resulting from reconnection and these loops later grow into the sigmoid. However, at the time of the EIS raster scan, the AIA 131 \AA{} loops have faded and the EIS data show that the box covers a set of loops which exhibit a red-shift (Figure \ref{fig:EIS_FIP}). Along the line-of-sight there will be a contribution from photons of arcade plasma that is not part of the flux rope as well as plasma in the flux rope itself. The blue box samples plasma that appears to be within the western leg of the rope. Photons from this plasma are blue-shifted, revealing that in this location there is a plasma flow toward the observer along the line-of-sight (Figure \ref{fig:131_Multi}d). We conclude that the blue box samples plasma within the leg of the flux rope because it is within the dimming region seen in EUV (marked DR in Figure \ref{fig:131_Multi}d and seen in Figure \ref{fig:Dimmings}a) which traces back to the hooked flare ribbon.

The analysis returned a FIP bias of 3.1 at the northern loops and 1.9 for the southern region. FIP biases of the order 1.0 are representative of photospheric plasma, whereas FIP biases of the order of 2-3 correspond to coronal plasma composition (for a review of the FIP effect including the interpretation of FIP biases, see  see \citealp{laming2015fip}). Therefore, these values both suggest plasma that is coronal in composition rather than photospheric. Uncertainties in the FIP bias factors are difficult to quantify, since errors in the radiometric calibration and atomic data are likely to be systematic in nature. Here we use the standard deviation from a distribution of values calculated from 1000 Monte Carlo simulations, where the intensities are randomly perturbed within the calibration error. This produces an uncertainty of $\approx$0.3, which is much less than the difference between photospheric abundances (FIP bias $\approx$1) and the range of coronal abundances ($\approx$2-3).

\subsection{Coronal Radio Observations} \label{radio}
 
 \begin{figure} 
\centering
\includegraphics[width=0.78\textwidth,trim=5 5 0 5, clip]{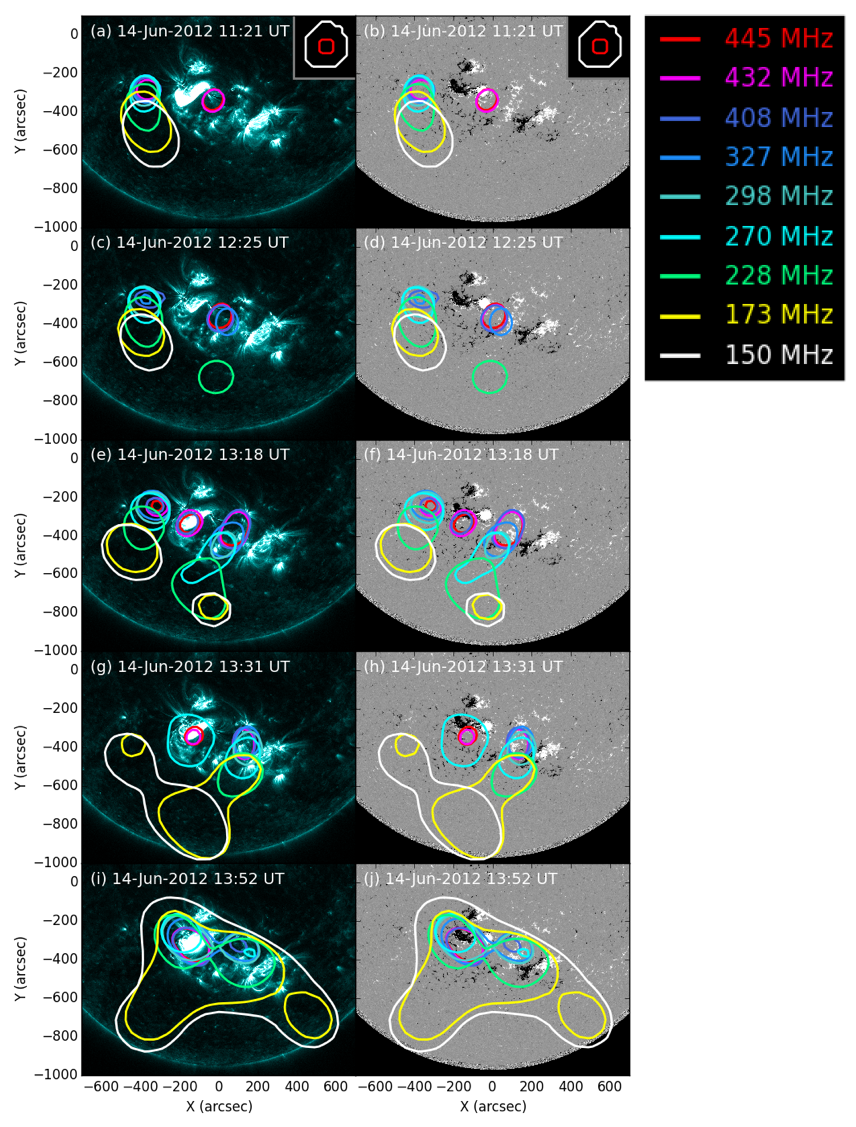}
 \caption{Contours representing locations of radio emission at different frequencies around NOAA active region 11504. Contours are drawn at 50\% of the maximum radio intensity at each frequency and time. The beam sizes of NRH at 150 MHz (largest) and 445 MHz (smallest) at 11:21 UT on 14 June 2012 are indicated in the top-right corners of panels a and b, to show that all sources are resolved. The contours are overlayed on 131 \AA{} images taken by AIA (left) and line-of-sight magnetograms taken by HMI (right) at the closest available times to the radio data. (a/b) Approximately two hours before the eruption, ongoing radio emission at all observed frequencies is present to the east of the active region. High-frequency emission is seen to the south-west of the positive sunspot for 20 minutes. (c/d) The high-frequency emission continues to the south-west of the positive sunspot, but slightly further west than in panels a/b. Some 228 MHz emission is also observed further south. (e/f) High-frequency radio emission is observed in the core of the active region and a range of frequencies are observed to the south-west. The high-frequency western radio sources appear even further to the west of the active region than in the previous panels and the low-frequency eastern radio sources appear deflected away from the active region. (g/h) As the eruption is beginning, the eastern emission ceases, lower frequencies are observed in the core, and even lower frequencies are observed centrally to the south. (i/j) During/after the eruption, large areas of low frequencies are observed to the south-west and south-east. 
(An animated version of this figure is available in the online journal.)}
 \label{fig:Radio}
 \end{figure}

Solar radio emission originates from the acceleration of particles, which can occur during magnetic reconnection. The coherent mechanism that can cause radio emission from the accelerated particles results in lower frequencies of radio emission from source regions of lower-density plasma, and higher frequencies from higher density plasma (see \textit{e.g.}, \citealp{mclean1985radiophysics,pick2008sixty,reid2014review}, as reviews). Observing the Sun at a number of different frequencies enables the probing of radio emission from different plasma densities. Since plasma density generally varies with altitude in the solar atmosphere (with lower plasma density at larger heights), radio emission at lower frequencies generally originates from higher altitudes, whilst high-frequency emission generally originates lower down. However, this is not strictly true because plasma density also varies between different structures, \textit{e.g.}, coronal loops and coronal holes.

Radio observations taken at nine frequencies between 150 MHz and 445 MHz from 10:00 UT on 14 June 2012 show strong emission across all frequencies to the east of the active region until 13:30 UT (Figure \ref{fig:Radio}a--f). The emission at different frequencies appears to form a column, suggesting radio emission from different coronal heights. The sources of emission appear progressively further to the south as frequency decreases, likely because the emission at lower frequencies originates from higher altitudes, and is therefore subject to stronger projection effects. This is a type I noise storm, which corresponds to acceleration of electrons in the corona over a continuous period. During this same time period, the \textit{Wind} spacecraft observed many successive individual type III radio bursts at lower frequencies (from 14 MHz down to 0.1 MHz), indicative of a type III noise storm. Frequencies of 0.1 MHz imply that the accelerated electrons generated coherent radio emission from rarefied plasma as far out as 0.3 AU (\textit{e.g.}, \citealp{leblanc1998tracing,mann1999heliospheric}). The type III bursts therefore indicate that the accelerated electrons had access to magnetic fields which extended out of the corona, and could subsequently travel through interplanetary space.

\begin{figure} 
 \centerline{\includegraphics[width=0.7\textwidth,clip=]{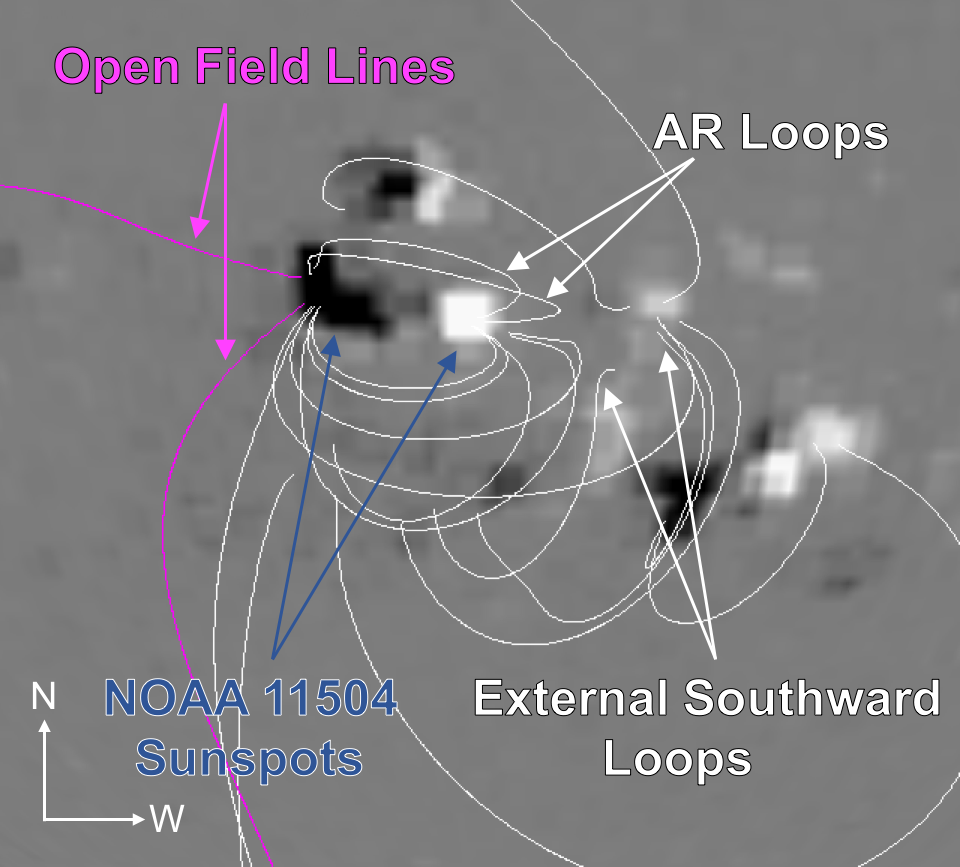}}
 \caption{A potential field source surface (PFSS) extrapolation of the solar magnetic field at 12:04 UT on 14 June 2012 produced using the pfss package of \texttt{SolarSoft}. Field lines are plotted on an image of the photospheric magnetic field and extend from positive polarities (white) towards negative (black). The sunspots that comprise NOAA active region 11504 are indicated with dark blue arrows. Closed field lines are shown in white and field lines that are open to the heliosphere are magenta. The extrapolated field lines originate from a height of 1.15 R$_{\odot}$ with a uniform grid spacing and the viewing perspective is from that of Earth. The extrapolation suggests the presence of open field lines to the east of NOAA active region 11504, active region loops that initially curve north-westward from the positive sunspot (AR loops) and loops that originate from an area of positive magnetic polarity to the west of the active region and connect to negative field to the south of the sunspots (external southward loops).}
 \label{fig:PFSS}
 \end{figure}

To further investigate the coronal magnetic field, we performed a potential field source surface (PFSS) extrapolation using line-of-sight photospheric magnetic field data at 12:04 UT on 14 June 2012 (see Figure \ref{fig:PFSS}). Although PFSS extrapolations have limitations, since they necessarily produce a potential field, they are useful in probing the global magnetic field configuration. 
This extrapolation suggests the presence of a region of magnetic field that is open to the heliosphere prior to the eruption. This open field is situated to the east of the active region and corresponds to the location of the radio noise storm seen in the NRH images. Open field allows particles accelerated at different altitudes (plasma densities) to escape into the heliosphere, which is commensurate with the observation of radio emission at all of the NRH frequencies.

A faint radio source is also present at the west of the active region in the higher NRH frequencies (\textit{e.g.}, 445 MHz, see Figure \ref{fig:Radio}a/b). From 11:20 UT (around the same time as the C5.0 soft X-ray flare and the initial brightening of loops that later comprise the western portion of the sigmoid), the western radio source brightens, indicating increased particle acceleration. Higher radio frequency type I emission corresponds to higher density plasma, and so the increased particle acceleration is likely occurring low-down in the solar atmosphere. The eastern noise storm remains present and type III bursts continue to appear in the interplanetary data, but at 12:25 UT, the lower frequency sources begin to move further eastward, giving the appearance that the column is being deflected away from the active region. Also around 12:25 UT, relatively low-frequency radio emission is briefly observed centrally to the south of the active region before reappearing and remaining from 13:08 UT. With projection effects, this could represent high-altitude emission above the centre of the active region, and the lack of spread in frequency implies that we are observing emission originating only from a certain height in the solar atmosphere. 

Beginning at around 13:18 UT, a new radio source appears centred on the active region at frequencies above 327 MHz (Figure \ref{fig:Radio}e/f). We interpret this emission to originate from the core of the active region because it is localised to the higher frequencies. From this time, the eastern noise storm begins to disappear, completely vanishing around 13:30 UT. Meanwhile, the western signatures extend into lower frequencies, developing to show a column of emission from different frequencies (Figure \ref{fig:Radio}g/h). The lower-frequency sources extend southward until they reach the same location as the central high-altitude emission. The sources of this emission move outwards to the west of the active region (see Figure \ref{fig:Radio}c-h).

After 13:30 UT, the central high-altitude radio emission begins to move further southward, consistent with a moving type IV radio burst (accelerated electrons within plasma that is erupting into interplanetary space, \textit{e.g.}, \citealp{bain2014radio} --- see Figure \ref{fig:Radio}g/h). The moving type IV source is observed to move from (-125$''$,-875$''$) to (-225$''$,-1050$''$) in 8 minutes, and so, assuming radial expansion from the active region latitude of 17$^{\circ}$, the speed of the source is $\approx$ 1040 km s$^{-1}$. This speed is of the same order as the 980 km s$^{-1}$ plane-of-sky CME speed estimated by LASCO.

After the eruption, around 13:52 UT, low-frequency (high altitude) radio signatures are observed over the active region core and western external polarities, as well as to the south-east and south-west of the active region (Figure \ref{fig:Radio}i/j). The southern sources at 150 MHz move away from the active region to the south-east and south-west before fading after 15 minutes. Finally, from 14:00 UT onwards, emission continues in all frequencies to the east and west of the active region, and then only from the eastern location after 14:13 UT.

\section{Discussion} \label{Discussion}

Here we bring together the wide range of observations presented in the previous sections. We aim to build a coherent interpretation of how the magnetic field in NOAA active region 11504 evolved in the hours leading up to the eruption on 14 June 2012. The studied data cover plasma emitting across a range of temperatures and altitudes from the photosphere to the corona.

\begin{figure} 
 \centerline{\includegraphics[width=1.0\textwidth,clip=]{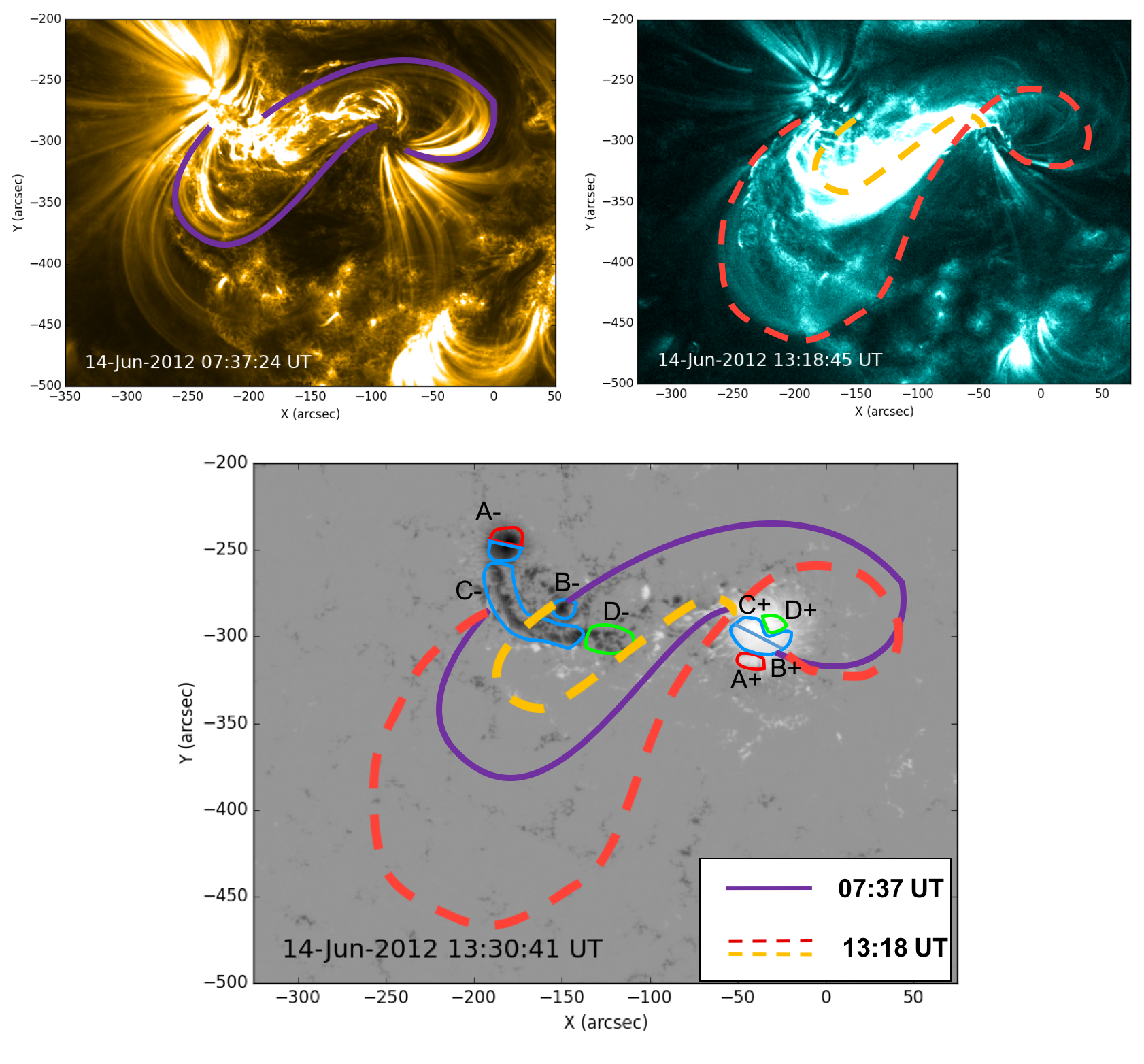}}
 \caption{Schematic to illustrate the formation of the sigmoid and arcade. Initially, sheared loops (seen here in 171 \AA{} and indicated with purple lines) connect from B+ to B- and from C+ to C-, using the nomenclature of Figure \ref{fig:HMI_multi}. Later, a flare arcade and a sigmoid (shown in dashed gold and red respectively) are observed. The sigmoid connects from B+ to C-, and the arcade footpoints are at C+ and B-. The sigmoid footpoints are consistent with the footpoint locations inferred from the EUV dimmings shown in Figure \ref{fig:Dimmings}. The transition from sheared loops to an arcade and sigmoid is consistent with tether-cutting-like reconnection, although it differs from that described by \citet{moore2001onset} in that the flux rope that forms (evidenced by the sigmoid) is stable for at least 2 hours. The background image of the lower panel corresponds to the line-of-sight magnetic field distribution shown in Figure \ref{fig:HMI_multi}g with the annotated coloured contours of Figure \ref{fig:HMI_multi}h that show which areas of the flux distribution emerge at the same time and are therefore connected.}
 \label{fig:tether}
 \end{figure}

The AIA observations strongly support an interpretation that part of the magnetic field in NOAA active region 11504 transforms from a sheared arcade to a flux rope before the CME occurs. This is illustrated by a cartoon shown in Figure \ref{fig:tether}, where two sets of sheared loops are indicated on a 171 \AA{} image at 07:37 UT on 14 June 2012. Then, by 13:18 UT, a full S-shaped (sigmoidal) emission structure is observed, most prominently in the AIA 131 \AA{} channel along with a flare arcade. The earliest signature of the forming sigmoid is seen at 11:20 UT (see Figure \ref{fig:131_Multi}b), although this then fades and becomes visible again at around 12:20 UT. The sheared loops are connected to the areas of magnetic flux that emerge between 11-13 June, labelled B$\pm$ and C$\pm$ in Figure \ref{fig:HMI_multi}, and the footpoints of the subsequently formed sigmoid and flare arcade are in approximately the same locations. A process similar to the tether-cutting magnetic reconnection described by \citet{moore2001onset} could explain this transition from sheared loops to a sigmoid and flare arcade with conserved footpoints. However, the reconnection did not apparently lead to the immediate ejection of the flux rope as in the \citet{moore2001onset} scenario.

The sigmoid is difficult to see in its central section due to emission from the central arcade of the active region (which is below the sigmoid but along the same sight-line), however the large extent of the S-shaped structure allows it to be identified. The magnetic tongues in the emerging photospheric magnetic field in the active region indicate that the sigmoid forms in a flux region which is of right-handed chirality, consistent with the weak trend for active region magnetic fields in the southern hemisphere. In addition, the sense of the S shape (forward-S) also indicates right-handed chirality \citep{pevtsov1997subphotospheric}. A stable forward-S sigmoid in positive chirality field is consistent with the presence of helical field lines that have a dip in their central section \citep{toeroek2010writhe}, which we interpret as a flux rope with at least one turn in the field lines (as per our definition in Section \ref{Intro}). Previous work has shown that sigmoidal emission can be produced by plasma trapped on S-shaped field lines that pass through a quasi-separatrix layer underneath a flux rope \citep{titov1999basic}. Furthermore, the hooked shape of the western flare ribbon is considered a signature of energy deposition along field lines at the periphery of a flux rope \citep{demoulin1996three,janvier2014electric}. Our interpretation of the presence of a flux rope in the active region by the time of the CME is matched by the \textit{in situ} data, which find the interplanetary CME to contain a flux rope with a matching chirality and axial orientation to that observed at the Sun \citep{palmerio2017insitu}.

The location of the footpoints of the erupting flux rope are determined using EUV dimmings and the curved ends of the flare ribbons. The eastern EUV dimming is located over a patch of negative magnetic polarity to the south of the pre-existing eastern sunspot in the extended magnetic field region. The western EUV dimming and flare ribbon indicates that the western flux rope footpoint is located in the southern penumbra of the pre-existing positive sunspot. These footpoint locations are displayed in Figure \ref{fig:Dimmings}.

The sigmoidal structure that first partially appears in the 131 \AA{} images around 11:20 UT is co-temporal with the brightening of high-frequency (low-altitude) radio emission produced on the western side of the active region. The sigmoidal emission fades and then appears again at around 12:20 UT on 14 June 2012 --- just over an hour before the eruption and $\approx$30 minutes before the slow rise in the X-ray flux began. Whilst the 131 \AA{} channel of AIA does include contributions from cooler plasma at $\approx$0.4 MK, images taken in this channel are largely representative of coronal plasma temperatures $\approx$10 MK during flares or high temperature sigmoid formation. This indicates that the sigmoidal structure contains hot plasma. Both appearances of the S-shaped emission structure are associated with increases in soft X-ray emission detected by GOES and an enhancement in intensity and extent of the central arcade of the active region as seen by AIA. We interpret these observations as an indication that increased magnetic reconnection occurs in two phases and contributes to building a flux rope in the corona. The first phase of reconnection, around 11:20 UT, heats plasma and accelerates electrons, which lead to the radio emission at the western end of the flux rope. The second phase of magnetic reconnection, around 12:25 UT, then further builds the flux rope structure. The flare arcade brightens and grows simultaneously with the overlying sigmoid, suggesting they both develop as a result of the same episode of magnetic reconnection, and therefore that the flux rope builds during the times the flare arcade is seen to brighten. Furthermore, little photospheric flux cancellation is observed, there are no photospheric inverse crossings seen along the central section of the polarity inversion line, and the sigmoid formation is best observed in the 131 \AA{} channel of AIA ($\approx$10 MK; representative of flaring plasma in the corona). Together, these observations suggest that the flux rope does not form via reconnection in the photosphere or chromosphere via the flux cancellation formation process described by \citet{van1989formation}, but rather the flux rope forms at a higher altitude by reconnection in the corona.

The interpretation that magnetic reconnection in the corona builds the flux rope is further supported by the plasma composition in the flux rope. In a previous study, \citet{baker2013plasma} used plasma composition to investigate the altitude at which magnetic reconnection occurs. In that work, a sigmoidal active region was observed with a core channel of plasma that had a FIP-bias of order 1.0. This signified that the channel contained plasma of a photospheric composition. The conclusion of \citet{baker2013plasma} was that a flux rope formed via magnetic reconnection in the photosphere associated with flux cancellation along the polarity inversion line of an active region. However, in our study, a photospheric plasma composition is not observed in the sigmoidal structure. Instead, the composition analysis indicates that plasma in the initial sigmoid loops that begin to brighten around 11:20 UT and plasma within western leg of the flux rope is of coronal composition, with FIP biases of 3.1 and 1.9 respectively, with errors $\approx$0.3. This supports our interpretation that the flux rope forms in the corona. Coronal flux ropes, formed via coronal reconnection, have previously been observed in other works as the appearance of diffuse emission structures in the hottest AIA channels above brightening flare arcades (\textit{e.g.}, \citealp{reeves2011atmospheric,zhang2012observation,patsourakos2013direct}). In these cases, flux ropes were viewed along their axes when the structure was at the limb. Here, we observe the sigmoid close to disc centre, viewing the flux rope from above and allowing the photospheric magnetic field evolution to also be studied. So, now we turn to look at what role the photospheric flows might have played in the evolution of the magnetic field in the corona.

As discussed previously, NOAA active region 11504 was in its emergence phase at the time the CME studied here erupted. The flux emergence proceeded in stages with different bipoles emerging in sequence. On 11 June, there were already two small sunspots present, and the emergence of further bipoles, some of which coalesced with the pre-existing field, developed these sunspots. The HMI data show that the photospheric horizontal magnetic field component is sheared in the locations of the emerged flux, \textit{i.e.}, to the north of the positive sunspot and in the negative elongated flux region, which manifests as strong concentrations of vertical electric current. There are domains of positive emerged flux that remain distinct from the main positive sunspot (leading spot) as well as a region of negative magnetic flux that forms an elongated structure to the south of the main negative sunspot (trailing spot). As the flux emergence proceeds, there are rotational and shearing motions between the bipoles. The positive magnetic flux that emerges between 13 June and 15 June moves westwards towards the pre-existing positive sunspot and then moves clockwise around the pre-existing sunspot umbra as separate umbrae. This suggests that, in the days leading up to the eruption, the coronal magnetic field emanating from this strongly moving and shearing positive region wraps around the magnetic field of the pre-existing sunspot. \citet{yan2012sunspot} also observed one umbra rotating around another, and associated the anti-clockwise rotation they observed with the formation of a left-handed sigmoid (whereas here we see clockwise rotation and the formation of a right-handed sigmoid). \citet{yan2012sunspot} concluded that the twisting of magnetic field from the rotational motion led to a non-potential field configuration higher up in the atmosphere. Similarly, we conclude that the clockwise wrapping of twisted and sheared emerging magnetic flux facilitates the formation of the flux rope in this study.

It is the collision between the two sheared domains of newly emerged flux that can facilitate the magnetic reconnection in the corona that subsequently builds the sigmoidal structure (and inferred flux rope). As discussed previously, this magnetic reconnection is likely to be of tether-cutting type \citep{moore1980filament,moore2001onset} that forms a flux rope through episodes of magnetic reconnection between two sets of sheared loops (illustrated in Figure \ref{fig:tether}), but which here initially produces a stable flux rope that only later erupts rather than runaway reconnection that both builds the flux rope and facilitates the eruption. Numerical studies have shown that a thin current layer can be formed in an arcade that is subject to shearing, and that in the presence of resistivity, magnetic reconnection can proceed in this current layer \citep{mikic1994disruption,roumeliotis1994numerical,amari1996plasmoid}. We propose that the shearing and rotational motion of the positive magnetic field elements around each other in NOAA active region 11504 leads to magnetic reconnection in the sheared arcade and the formation of a magnetic flux rope in the corona.

The photospheric motions are likely to also play a role in the occurrence of the CME beyond the formation of the flux rope itself. The shearing and strong motions of the emerging flux (especially in the positive field regions) could inflate the global magnetic field of the active region, eventually bringing the configuration to the point where the flux rope experiences a loss of stability or equilibrium \citep{torok2013initiation}. The rotational motion seen here in the western sunspot of NOAA active region 11504 could have driven the twisting of field rooted in and around the sunspot, causing it to rise and inflate. Furthermore, other field rooted in the vicinity that overlies the forming flux rope may also twist and inflate, up to the point of becoming unable to contain the underlying flux rope due to decreased tension. The eastward deflection of the radio sources in the eastern column of type I radio emission and the westward motion of the western sources mentioned in Section \ref{radio} and seen in Figure \ref{fig:Radio} suggests that the radio sources are moving away from the active region before the eruption. This is consistent with an expanding magnetic field configuration.

The increase in radio emission and the production of emission at lower frequencies to the western side of the active region in the lead-up to the eruption suggests that magnetic reconnection is ongoing, with electrons being accelerated at increasing altitudes, perhaps a consequence of the magnetic field of the active region and the flux rope inflating. This takes place during the slow-rise phase of the eruption. Furthermore, high-altitude radio emission is observed directly above the active region from 13:18 UT, which is also during the slow-rise phase. The high-altitude emission source appears stationary at first and indicates the acceleration of electrons produced by reconnection as the expanding active region field pushes against and reconnects with the surrounding field above the active region. After 13:30 UT, the eruption moves into the fast-rise phase, the flux rope escapes, and more impulsive flare reconnection sets in. This is consistent with previous work that has shown that the slow-rise and fast-rise phases are common to CMEs \citep{zhang2006statistical}. In this study, the first phase (the slow-rise phase) lasts for around 30 minutes and shows a steady increase in soft X-ray flux which is co-temporal with the growth and expansion of the sigmoidal loops in 131 \AA{} (and therefore the flux rope) and the radio emission. The second phase is a sharp rise phase that occurs during the eruption of the sigmoid that involves more impulsive flare reconnection below and above the flux rope. The central high-altitude radio emission that is stationary during the slow-rise phase moves southward in the plane of the NRH images, and is interpreted as a type IV radio burst corresponding to the acceleration of energetic electrons within the plasma of the moving CME as it erupts.

It is interesting to note the timescale over which the flux rope appears to be stable. From the SDO/AIA observations and the appearance of the sigmoid, we infer that the flux rope forms at least 2 hours prior to its eruption. Few studies have been carried out to probe when flux ropes form prior to their eruption, and the timescales over which they are stable. Previous studies of flux ropes that form through magnetic reconnection in the corona, as in this event, suggest they might be stable for a few hours. For example, \citet{patsourakos2013direct} studied an event where the flux rope formed 7 hours prior to eruption and \citet{cheng2014formation} investigated the eruption of a flux rope that formed 2 hours beforehand. In contrast, a small survey by \citet{green2014observations} found that flux ropes that formed by photospheric magnetic reconnection associated with flux cancellation were stable in their active regions for between 5 and 14 hours. This is an area where more research should be conducted. Flux ropes formed by reconnection in the corona will be of the hyperbolic flux tube type, whereas flux ropes formed through flux cancellation may be of bald patch-type with their underside line-tied to the photosphere. The altitude and specific topology of the flux rope may affect the timescale over which it is stable, but no parametric numerical study of this has yet been conducted.

\section{Conclusions} \label{Conclusions}

In this study, we investigate the pre-eruptive configuration of a flux rope that is detected \textit{in situ} to determine whether there is a flux rope present before the onset of eruption at the Sun, or whether it forms at a later time. It is concluded that a flux rope forms via reconnection in the corona 2 hours before the onset of eruption. This is evidenced by a flux rope of coronal composition associated with a sigmoid that first partially flashes into view 2 hours prior to the eruption in 131 \AA{} images, before later reappearing, growing and erupting.

The CME occurs when NOAA active region 11504 is still in its emergence phase. Positive areas of emerging bipoles move towards and then clockwise around the leading positive sunspot, facilitating tether-cutting-type reconnection in the corona that forms a flux rope and an underlying flare arcade, but does not initially cause the flux rope to erupt. Photospheric motions may cause the inflation of the global active region magnetic field. Simulations have previously shown that vortical motion at the footpoints of magnetic loops can cause the loops to rise, at which point they may become unstable to eruption \citep{torok2013initiation}.

%

%
%

\begin{acks}
AWJ gives thanks to Yang Liu for providing HMI SHARP data for use in this work and to the M.S.S.L. solar group for their support during the first year of his PhD, during which time this paper was produced. 
AWJ, LMG, GV and LvDG acknowledge the support of the Leverhulme Trust Research Project Grant 2014-051. LMG also thanks the Royal Society for funding through their URF scheme. 
EP acknowledges the Doctoral Programme in Particle Physics and Universe Sciences (PAPU) at the University of Helsinki, the Finnish Doctoral Programme in Astronomy and Space Physics, the Magnus Ehrnrooth foundation, and the V\"ais\"al\"a Foundation for financial support. 
HASR acknowledges funding from an STFC consolidated grant ST/L000741/1. 
DB and LvDG are funded under STFC consolidated grant number ST/N000722/1. LvDG also acknowledges the Hungarian Research grant OTKA K-109276. 
The work of DHB was performed under contract to the Naval Research Laboratory and was funded by the NASA \textit{Hinode} program. 
EK acknowledges UH 3-year grant project 490162 and HELCATS project 400931. 
We thank the referee for their valuable, constructive feedback.
This research has made use of SunPy, an open-source and free community-developed solar data analysis package written in Python \citep{community2015sunpy}. 

\end{acks}

\medskip
\noindent
\textbf{Disclosure of Potential Conflicts of Interest}  The authors declare that they have no conflicts of interest.

%
%
%
%
%
%

\bibliographystyle{spr-mp-sola}
\bibliography{document} 

\end{article} 
\end{document}